\documentclass[12pt,a4paper]{article}
\pdfoutput=1
\usepackage{jheppub,amsmath,amssymb,color,youngtab}
\usepackage[nottoc]{tocbibind}
\newcommand{\bea}{\begin{eqnarray}}
\newcommand{\eea}{\end{eqnarray}}
\newcommand{\Tr}{{\rm Tr}}

\begin{document}
\title{Collective  Theory at Finite-$N$:Reduction of the Emergent Hilbert Space}
\author[a,b]{Robert de Mello Koch,}
\author[c,d]{Antal Jevicki,}
\author[e]{Garreth Kemp}
\author[b]{ and Anik Rudra,}
\affiliation[a]{School of Science, Huzhou University, Huzhou 313000, China}
\affiliation[b]{Mandelstam Institute for Theoretical Physics, School of Physics, University of the Witwatersrand, Private Bag 3, Wits 2050, South Africa}
\affiliation[c]{Department of Physics, Brown University,
182 Hope Street, Providence, RI 02912, United States}
\affiliation[d]{Brown Center for Theoretical Physics and Innovation, Brown University,
340 Brook Street, Providence, RI 02912, United States}
\affiliation[e]{Department of Physics, University of Johannesburg, Auckland Park, 2006, South Africa}
\date{October 2025}
\abstract{Continuing the formulation of finite $N$ Hilbert spaces in emergent theories we study in this work $S_N$ symmetric collective models . For the case of $N$ bosons in $d$ dimensions, which map to  matrix models with commuting matrices, we describe a complete algorithm and give a detailed case study reproducing the expected primaries and determining secondary invariants at each bidegree (a Hironaka decomposition). 
The method is based on null spaces (of the full collective theory) which are seen to yield all the independent trace relations, reducing the construction to linear algebra.
 As a stringent check, of our algorithm, we have verified that the system of invariants generates a subset of gauge invariant operators with no redundancies. This results in a reduction of the  Hilbert space, in particular the gauge invariant secondary invariants realize an emergent Fock space with  finite-$N$ occupation-numbers.}
\maketitle

\section{Introduction}
Collective field theory is generally formulated on the space of gauge invariant (single-trace) operators, with $N$ appearing as a coupling constant of the theory. From an algebraic standpoint, this is indeed the simplest possible structure, correct at large $N$ (and even perturbatively in $1/N$). At finite $N$, however, well known trace relations intervene and one needs a more structured description. Recently, we have investigated this structure to some degree. It has been demonstrated, \cite{deMelloKoch:2025ngs}: that at finite $N$, a finite--$N$ ring of invariants remains operational, taking the form of a \emph{Hironaka decomposition}. Concretely, there exists a set of homogeneous, algebraically independent \emph{primary} invariants
\[
\{P_1,\dots,P_r\}\,,
\]
such that the full ring is a free, finitely generated module over the polynomial subring $\mathbb{C}[P_1,\dots,P_r]$. The additional generators, the \emph{secondary} invariants
\[
\{S_1=1,S_2,\dots,S_s\}
\]
form a basis for the module. Concretely, the space of gauge invariant operators at finite $N$, ${\cal H}_N$, decomposes as
\begin{equation}
\label{eq:direct-sum}
{\cal H}_N \;\cong\; \bigoplus_{\gamma=1}^{s}\, \mathbb{C}[P_1,\dots,P_r]\; S_\gamma
\end{equation}
and products of secondaries reduce \emph{linearly} over the primaries,
\begin{equation}
\label{eq:lin-reduction}
S_\alpha S_\beta \;=\; \sum_{\gamma=1}^{s} f_{\alpha\beta}{}^{\gamma}(P_1,\dots,P_r)\,S_\gamma\,,
\qquad f_{\alpha\beta}{}^{\gamma}\in\mathbb{C}[P_1,\dots,P_r]\,.
\end{equation}

\paragraph{Interpretation.}

The above structure, of the finite $N$ ring of invariants, translates directly into the finite $N$ Hilbert space of the emergent theory. The primaries, which are a subset of collective excitations of the original theory, are in agreement with the \emph{perturbative} degrees of freedom of the  theory: they act freely and generate a Fock sub-space of excitations. For a $d$–matrix model the number of algebraically independent primaries is the Krull dimension,
\begin{equation}
\label{eq:Krull}
r \;=\; 1+(d-1)N^2\,,
\end{equation}
so the primary sector scales extensively with $N^2$ as expected for adjoint large--$N$ theories. The secondary invariants act at most linearly. Although one does not have a closed count of all secondaries, general considerations~\cite{deMelloKoch:2025qeq} imply their number grows as $\exp\!\big(c\,N^2\big)$ for some order 1 number $c>0$. This scaling signals that typical secondary states are built from $\mathcal{O}(N^2)$ fields; they are heavy enough to backreact and thus correspond to new spacetime geometries in the dual gravitational description.

There is, further, a distinguished subset of secondaries with dimension $\mathcal{O}(N)$ (rather than $\mathcal{O}(N^2)$). These are light enough not to backreact and have a natural interpretation as solitonic excitations, e.g.\ giant graviton branes. They are sparse but must be present among the secondaries if the spectrum is to capture all solitons. Conversely, one also has primaries of very low dimension, $\leq N$. This is because all single–trace operators with fewer than $N\!+\!1$ letters must appear in the generating set; yet only $\mathcal{O}(N^2)$ of them can be algebraically independent primaries. Thus the vast majority of the exponentially many single--trace structures (schematically $\mathcal{O}(e^{N})$ at fixed alphabet) must enter as \emph{secondaries}.

\paragraph{Phase Transition.} Matrix models are well-known to exhibit Hagedorn behavior at infinite $N$ \cite{Sundborg:1999ue,Aharony:2003sx} with a finite transition temperature. For vector type theories the critical temperature,
is of order $N$ and the relevance of finite $N$ constraints in this regard was proposed in~\cite{Shenker:2011zf}. At large but finite $N$ a Lee-Yang behaviour was exhibited~\cite{Kristensson:2020nly}. The Lee–Yang transition is seen directly from the zeros of the finite-$N$ partition function, which condense on arcs in the complex fugacity plane and pinch the real axis at the critical point. While the critical non-analyticity is controlled by the asymptotic growth (the denominator), the finite-$N$ distribution of zeros -- and thus the observable onset and sharpness of the transition -- depend sensitively on the secondary invariants: numerator zeros interfere with the group-integral structure to position and weight the Lee–Yang arcs before they coalesce in the limit. Thus the combinatorics of the secondary invariants imprints directly on finite-$N$ thermodynamics. Beyond locating the putative critical point, a precise accounting of the secondary invariants is essential for predicting the thermodynamics in $S_n$-invariant ensembles~\cite{Hanada:2020uvt}.

\paragraph{Finite–$N$ cutoffs and $q$–reducibility.}
In explicit computations~\cite{deMelloKoch:2025rkw} we find a striking pattern we term \emph{$q$–reducibility}: among the secondary invariants there is a distinguished set of short single–trace operators, \(\{s_a\}\). Their products \(\prod_a s_a^{m_a}\) appear in the set of secondaries, much like a Fock construction. Finite–\(N\) trace identities, however, force these towers to \emph{truncate}: 
\begin{equation}
\label{eq:qcut-multi}
\prod_a s_a^{m_a}
\;=\;
\sum_{\gamma}\,f_\gamma(P_1,\dots,P_r)\; S_\gamma,
\end{equation}
for some $f_\gamma\in \mathbb{C}[P_1,\dots,P_r]$ and for $m_a$ beyond a cut off $q_a$. This is \emph{$q$–reducibility}: beyond the cutoffs $\{q_a\}$ no new independent secondaries are produced; putative new invariants are redundant by virtue of the secondary relations (\ref{eq:lin-reduction}). In the usual large–\(N\) holography intuition~\cite{Maldacena:1997re,Gubser:1998bc,Witten:1998qj}, where each single–trace $s_a$ behaves as a Fock oscillator and the power \(m_a\) is an occupation number, \(q\)–reducibility is a \emph{finite–\(N\) occupation number cutoff} enforced by trace relations. In the dual gravitational language, multiparticle graviton states with occupancies exceeding the cutoff are not independent implying a significant truncation of the high-energy spectrum of the emergent theory.

\paragraph{Collective field theory as a dynamical setting.}
The discussion so far has been algebraic. A natural dynamical framework that realizes these structures is collective field theory~\cite{JS1}, which uses the invariants as the fundamental dynamical degrees of freedom. It is worth emphasizing that collective field theory features $1/N$ as the loop expansion parameter. The overcomplete collective description  correctly reproduces the perturbative (in the sense of large $N$) features of the theory .It is less well known, that it also applies non-perturbatively,
ie at finite $N$. Crucially, the finite–$N$ constraints commute with the collective Hamiltonian~\cite{Jevicki:1991yi}, so they can be imposed as operator equations to eliminate redundant variables and arrive at the finite–$N$ theory. For general matrix models, this  has  been carried out at low $N$ i.e. for $N=3,4$~\cite{deMelloKoch:2025rkw}: complete sets of primary and secondary invariants have been explicitly constructed i.e. the constraint equations are solved in full.

\paragraph{A prototypical toy model.}To establish that the above program works generally, for higher $N$, in this paper we consider a simple model that captures all essential features while avoiding noncommutativity issues: an $S_N$ (symmetric–group) model of $N$ bosons in $d$ spatial dimensions. Although the finite–$N$ constraints are most transparently phrased in matrix language~\cite{Domokos}, in this model the relevant matrices \emph{commute}, leading to a major simplification. For this example we can evaluate the Molien–Weyl generating functions to arbitrarily high order using efficient recursion relations. The primaries are known explicitly~\cite{Domokos}, and we introduce a new, purely linear–algebraic numerical algorithm to construct the secondary invariants.

\paragraph{Algorithmic determination of secondaries.}
It is established that single traces of length $\le N$ generate all gauge–invariant operators; this fixes the ambient generating set. The primary invariants within this set are known. Determining the secondaries is then equivalent to finding a \emph{free} module basis for the quotient obtained after deleting the primaries. Operationally, we assemble the candidate secondary invariants degree by degree, form the matrix obtained by evaluating this set at a number of numerical values, and compute its \emph{null space}. The null vectors are precisely the relations among invariants and solving them yields a free generating set which is the set of secondary invariants. In the end, the entire procedure reduces to elementary linear algebra. The resulting structure makes the finite–$N$ cutoff on Fock–space occupation numbers manifest, thereby providing an algebraic derivation of the gravitational redundancy discussed above.

The paper is organized as follows. In the next section we review those aspects of collective field theory that are relevant for this study. Concretely we show how the finite $N$ constraints can be written as a collection of mutually commuting operators that also commute with the Hamiltonian. In Section \ref{perminvariants} we review necessary facts about the relevant algebra of invariants. In particular we review efficient methods to compute the Molien-Weyl function which counts the invariants, the structure of the primary invariants and an explicit generating set. In Section~\ref{algorithm} we present the details of our secondary construction algorithm. The algorithm is illustrated in detail for the example with $d=2$ and $N=3$. We go on, in Section \ref{dis2story} to analyze $d=2$ systems at arbitrary $N$. We find that the complete set of secondary invariants can be organized into towers and the rules for the construction of each tower can be written down as a function of $N$. In Section \ref{coinvariantalgebra} we review the coinvariant algebra, which is the quotient of the algebra of invariants by the ideal generated by the primaries. The structure of the coinvariant algebra naturally explains the structure of irreducible and reducible secondary invariants. We discuss our results in Section \ref{discussion}. The Appendices collect some technical details about the Molien-Weyl functions as well as the numerical tests we have performed of the systems of invariants we have constructed.

\section{Collective Field Theory}

The configuration space of a system of $N$ bosons in $d$ dimensions is given by the $Nd$ coordinates $x^a_i$ with $a=1,...,d$ and $i=1,...,N$. The collective field theory description of this system is based on the equal time invariant variables
\bea
\phi(n_1,n_2,\cdots,n_d)&=&\sum_{i=1}^N (x^1_i)^{n_1}(x^2_i)^{n_2}\cdots (x^d_i)^{n_d}\label{collflds}
\eea
and their canonical conjugates
\bea
\pi (n_1,n_2,\cdots,n_d)&=&{1\over i}{\partial\over\partial \phi(n_1,n_2,\cdots,n_d)}
\eea
The trace relations can be written as polynomial equations in the invariant variables
\bea
\Gamma(\{\phi\})&=&0
\eea
Our main goal in this Section is to illustrate, with detailed examples, that these finite $N$ constraints can be used to define eigenstates of the Hamiltonian given by
\bea
\Psi_{\Gamma(\{\phi\})}=\Gamma(\{\phi\})\Psi_0(\phi)
\eea
where $\Psi_0(\phi)$ is the ground state of the system. The projection operator, that projects onto a given eigenstate, commutes with the Hamiltonian and setting this projector to zero is equivalent to setting the trace relation to zero. These projectors also commute with each other.  In this way, the trace relations are realized as mutually commuting constraints that commute with the Hamiltonian. This establishes the consistency of enforcing the finite $N$ constraints on the unconstrained collective field theory's Hilbert space in order to recover the finite $N$ theory.

Although our argument is example based, the fact that the collective Hamiltonian always commutes with the finite-$N$ trace relations reflects a deep result of Procesi~\cite{Procesi2}. The over complete collective field theory is formulated at the level of the free algebra. Procesi’s result, which supplies a formal inverse to the Cayley–Hamilton theorem~\cite{Procesi2}, implies that the algebra of $S_N$-invariant polynomial functions is obtained from the free trace algebra by imposing all trace relations. At finite N, trace relations express universal\footnote{Let $\mathbb{F}$ be a field and $A$ an associative $\mathbb{F}$-algebra. We say that $A$ is a \emph{polynomial identity} (PI) algebra if there exists a nonzero polynomial $f(x_1,\dots,x_m)\in F\langle x_1,\dots,x_m\rangle$ in the free associative (noncommutative) algebra such that $f(a_1,\dots,a_m)=0$ for all $a_1,\dots,a_m\in A$~\cite{PI}.} PI-constraints rather than model-dependent dynamics.

\subsection{Trace Relations}

We consider a system described by the $N$-body Hamiltonian 
\bea
H&=&{1\over 2}\sum_{i=1}^N\sum_{a=1}^d \Big(-{\partial\over\partial x_i^a}{\partial\over\partial x_i^a}+x_i^ax_i^a\Big)\label{bosonhamiltonian}
\eea
The collective fields are given by the complete set of $S_N$-invariant combinations of the coordinates given in (\ref{collflds}). For any fixed $N$ these invariants are over complete. To derive the relations between them, introduce the $d$ $N\times N$ matrices defined by
\bea
X^a&\equiv&\left[\begin{array}{ccccc}
     x^a_1&0 &0  &\cdots &0\\
     0    &x^a_2 &0 &\cdots &0\\
     0    &0     &x^a_3 &\cdots &0\\
     \vdots &\vdots &\vdots &\ddots &0\\
     0 &0 &0 &\cdots &x^a_N
\end{array}\right]\label{matsdefined}
\eea
The relations between the invariant variables now follow by anti symmetrizing the column indices in the expression\cite{Pr}
\bea
\sum_{\sigma\in S_{N+1}}{\rm sgn}(\sigma)(W_1)_{i_1i_{\sigma(1)}}(W_2)_{i_2 i_{\sigma(2)}}\cdots (W_{N+1})_{i_{N+1}i_{\sigma(N+1)}}=0\label{PermCayleyHamilton}
\eea
where ${\rm sgn}(\sigma)$ is the parity of $\sigma$, and where the $W_a$ are each any word constructed out of the $X^a$. This identity is true because antisymmetrizing $N+1$ indices that each take $N$ values always vanishes.

\subsection{$d=1$}

In $d=1$ the invariants are labelled by a single integer $\phi(n)$. Changing variables from $x_i^a$ to the invariant variables $\phi(n)$ we obtain the following collective Hamiltonian
\bea
H&=&H_2+{1\over 2}\phi(2)\label{collectiveH}
\eea
where
\bea
H_2&=&-{1\over 2}\sum _{n=1}^\infty \sum _{m=1}^\infty m n \, \phi (m+n-2) \frac{\partial }{\partial \phi (m)}\frac{\partial}{\partial \phi (n)}\cr\cr
&&-{1\over 2}\sum _{n=2}^\infty n (n-1) \phi (n-2) \frac{\partial}{\partial \phi (n)}
\eea
This Hamiltonian is not manifestly Hermitian, signaling a non-trivial measure associated to the change of variables. Accounting for the measure we could obtain a manifestly Hermitian Hamiltonian~\cite{JS1}, but the above expression is perfectly suitable for our analysis. For the purpose of illustration, consider the first three trace relations following from (\ref{PermCayleyHamilton})
\bea
\Gamma_1(\phi)&=&\phi (1)^2-\phi (2)\cr\cr
\Gamma_2(\phi)&=&\phi (1)^3-3 \phi (2) \phi (1)+2 \phi (3)\cr\cr
\Gamma_3(\phi)&=&\phi (1)^4-6 \phi (2) \phi (1)^2+8 \phi (3) \phi (1)+3 \phi (2)^2-6 \phi (4)
\eea
Assign the product $\phi(n_1)\phi(n_2)\cdots\phi(n_k)$ the degree $n_1+n_2+\cdots+n_k$. We can see that $\Gamma_1$ is degree 2, $\Gamma_2$ is degree 3 and $\Gamma_3$ is degree 4. Every constraint $\Gamma_a$ has a definite degree $d_a$.

The ground state wave function is given by $\Psi_0(\phi)=e^{-{\phi(2)\over 2}}$. It is completely straightforward to verify that
\bea
\Big(H_2+{1\over 2}\phi(2)\Big)\Psi_0(\phi)&=&{N\over 2}\Psi_0(\phi)
\eea
A very similar computation shows that
\bea
\Big(H_2+{1\over 2}\phi(2)\Big)\Gamma_a(\phi)\Psi_0(\phi)&=&\left({N\over 2}+d_a\right)\Gamma_a(\phi)\Psi_0(\phi)
\eea
where $d_a$ is the degree of the constraint $\Gamma_a$. This demonstrates that every constraint is indeed associated to an energy eigenstate.

\subsection{$d=2$}

In this section we will consider the model with $d=2$ which involves the pair of matrices $X^1$ and $X^2$. Invariants are labelled by a pair of integers and the collective Hamiltonian is now given by 
\bea
H&=&H_2+{1\over 2}\big(\phi(2,0)+\phi(0,2)\big)\label{collectiveH2}
\eea
where
\bea
H_2&=&-{1\over 2}\sum _{\substack{n,m=0\\{\rm exclude}\,\, n=m=0}}^\infty\Big((n-1) n \phi (n-2,m)+(m-1) m \phi (n,m-2)\Big) \frac{\partial}{\partial \phi (n,m)}\cr\cr
&&-{1\over 2}\sum _{\substack{n_1,m_1=0\\{\rm exclude}\,\,n_1=m_1=0}}^\infty \sum _{\substack{n_2,m_2=0\\{\rm exclude}\,\, n_2=m_2=0}}^\infty \Big(n_1 n_2 \phi (n_1+n_2-2,m_1+m_2)\cr\cr
&&+m_1 m_2 \phi (n_1+n_2,m_1+m_2-2)\Big) \frac{\partial }{\partial \phi (n_2,m_2)}\frac{\partial}{\partial \phi (n_1,m_1)}
\eea
The first few trace relations are
\bea
\Gamma_1(\phi)&=&\phi (1,0)^2-\phi (2,0)\qquad\qquad\qquad\qquad\qquad 
\Gamma_2(\phi)\,\,=\,\,\phi (0,1)^2-\phi (0,2)\cr\cr
\Gamma_3(\phi)&=&\phi (0,1) \phi (1,0)-\phi (1,1)\cr\cr
\Gamma_4(\phi)&=&\phi (1,0)^3-3 \phi (2,0) \phi (1,0)+2 \phi (3,0)\qquad
\Gamma_5(\phi)\,\,=\,\,\phi (0,1)^3-3 \phi (0,2) \phi (0,1)+2 \phi (0,3)\cr\cr
\Gamma_6(\phi)&=&\phi (0,1) \phi (1,0)^2-2 \phi (1,1) \phi (1,0)-\phi (2,0) \phi (0,1)+2 \phi (2,1)
\eea
The relations $\Gamma_1,\Gamma_2$ and $\Gamma_3$ have degree 2, while $\Gamma_4,\Gamma_5$ and $\Gamma_6$ all have degree 3. 

The ground state wave function is given by $\Psi_0(\phi)=e^{-{1\over 2}(\phi(2,0)+\phi(0,2))}$. We now have
\bea
\Big(H_2+{1\over 2}\phi(2)\Big)\Psi_0(\phi)&=&N\Psi_0(\phi)
\eea
as well as
\bea
\Big(H_2+{1\over 2}\big(\phi(2,0)+\phi(0,2)\big)\Big)\Gamma_a(\phi)\Psi_0(\phi)&=&\left(N+d_a\right)\Gamma_a(\phi)\Psi_0(\phi)
\eea
where $d_a$ is the degree of the constraint $\Gamma_a$. This again demonstrates that every constraint is indeed associated to an energy eigenstate. The conclusion is easy to demonstrate for $d>2$ with simple computation.

\section{$S_N$ invariants}\label{perminvariants}

In this section we review the background material that underlies the construction algorithm presented in Section~\ref{algorithm}. In Section~\ref{genandprim} we exhibit a concrete generating set of $S_N$–invariant operators that generates the full ring of invariants. Within this set we isolate the family of \emph{primary} invariants. Section~\ref{mwproperties} then reviews the Molien–Weyl integral for the Hilbert series $H(t)$ of the invariant ring and records two structural consequences that we will exploit repeatedly: (i) the denominator, $\prod_{i=1}^{r}(1-t^{\deg P_i})$, encodes the degrees of the primaries; (ii) the numerator, $\sum_{\gamma} t^{\deg S_\gamma}$, counts the \emph{secondary} invariants. A key feature for our purposes is that the numerator is \emph{palindromic}. This plays an important role in our construction algorithm of Section~\ref{algorithm}. A good general reference for this section is~\cite{Sturmfels}.

\subsection{Generating invariants and primary invariants}\label{genandprim}

This Section constructs an explicit generating set for the invariant algebra by exploiting the trace identities familiar from matrix models. As explained above, we promote the particle coordinates to diagonal matrices, recasting the problem in a trace-algebra framework where the requisite relations can be organized and solved systematically~\cite{Pr}. The invariant algebra $\mathbb{C}[V^{Nd}]^{S_N}$ is generated by arbitrary polynomials in trace words constructed from the matrices $X^a$. These invariants can be graded by their degree in each $X^a$. Since the $X^a$ are simultaneously diagonal, they commute, and therefore any two words with the same multidegree (the same degree in each $X^a$) coincide; the corresponding invariants are identical. We define the total degree as the sum of the individual $a$-degrees, which simply counts the total number of matrices appearing in the invariant. The resulting trace relations are the standard trace identities for $N\times N$ matrices.

Using this framework, the algebra $\mathbb{C}[V^{Nd}]^{S_N}$ is generated by single-trace operators of degree at most $N$ and the only relations among these generators are the usual trace identities. The proof of this fact is simple~\cite{Domokos}: given a monomial of degree $\ge N+1$, write it as a product $w_1,\cdots,w_{N+1}$ of $N+1$ non-empty monomials. The trace relation (\ref{PermCayleyHamilton}) implies that $\Tr (w_1 \cdots w_{N+1})$ can be expressed in terms of traces of monomials of strictly smaller degree. This obviously implies that traces of degree $\le N$ generate the algebra of invariants. Consequently, the algebra $\mathbb{C}[V^{Nd}]^{S_N}$ is generated by
\bea
I_{n_1,n_2,\cdots,n_d}&=&\Tr( (X^1)^{n_1}(X^2)^{n_2}\cdots (X^d)^{n_d})\qquad \sum_{i=1}^d n_i\le N \label{allgens}
\eea
This generating set has a single invariant of each multidegree.

For a finite group acting on a polynomial ring over a field of characteristic zero, the Hochster–Roberts theorem ensures that the invariant ring is Cohen–Macaulay~\cite{hrtheorem}. Consequently, it admits a Hironaka decomposition: the algebra of invariants is a free module over the subring generated by the primary invariants, with a basis given by the secondary invariants. Our task is to extract the primary and secondary invariants from the generating set described above. The explicit form of the primary invariants is known~\cite{Domokos}. There are $dN$ primary invariants, given by
\bea
\label{eq:Primaries}
P^a_n&=&\sum_{i=1}^N (x_i^a)^n\qquad a=1,\cdots,d\qquad n=1,\cdots,N\label{primaryinvariants}
\eea 

To identify the secondary invariants, we remove the primary invariants (\ref{primaryinvariants}) from the full generating set (\ref{allgens}). The single trace operators that remain give the single trace secondary invariants. Products of these single trace operators are candidate secondary invariants. By looking at the numerator of the Molien-Weyl function we can read off the degree of all secondary invariants. Using the single trace secondary invariants we easily generate the possible candidates for a secondary invariant of any given degree. To test which of these is the correct choice we appeal to the trace relations. Our numerical algorithm reduces this to a straightforward linear algebra problem, which can be solved efficiently by numerical methods, as described in the next section.

There is an efficient way to evaluate the Hilbert series computed using the Molien-Weyl function. We grade by dimension so that the Molien-Weyl partition function is written as a function of $d$ variables $t_a$ and of $N$. It takes the form
\bea
\tilde{Z}_N(t_a)&=&\frac{N_N(t_a)}{\prod_{a=1}^d\prod_{i=1}^N (1-t_a^i)}\,\,=\,\,\frac{1+\sum_{n_1,\cdots,n_d}c_{n_1\cdots n_d}(t_1)^{n_1}\cdots(t_d)^{n_d}}{\prod_{a=1}^d\prod_{i=1}^N (1-t_a^i)}
\eea
where the denominator reflects the primary invariants (\ref{primaryinvariants}) and the coefficient $c_{n_1\cdots n_d}$ counts the number of secondary invariants of multidegree $(n_1,n_2,\cdots,n_d)$ in the matrices $X^a$. To determine $\tilde{Z}_N(t_a)$ we need to determine the numerator $N_N(t_a)$. Start with $d=2$. There is a known generating function \cite{bimahonian} (see also \cite{Rosele}) for $\tilde{Z}_N(t_1,t_2)$, given by 
\bea
F(u)\,\equiv\,\prod_{i,j\ge 0}\frac{1}{1-t_1^i t_2^j u}
&=& \sum_{N\ge 0} u^N \,\tilde{Z}_N(t_1,t_2)\cr\cr
&=& \sum_{N\ge 0} u^N \,\frac{N_N(t_1,t_2)}{\prod_{i=1}^N (1-t_1^i)(1-t_2^i)}
\eea
where $N_N(t_1,t_2)$ is the numerator polynomial. Take the logarithm of $F(u)$ which converts the product over $i,j$ into a sum. Power series expand in $u$ and then sum over $i$ and $j$ to find
\bea
\log F(u) %&=& -\sum_{i,j\ge 0} \log\!\bigl(1-t_1^i t_2^j u\bigr)\cr\cr
%&=& \sum_{i,j\ge 0}\sum_{k\ge 1}\frac{(t_1^i t_2^j u)^k}{k}\cr\cr
%&=& \sum_{k\ge 1}\frac{u^k}{k}\left(\sum_{i\ge 0} t_1^{ik}\right)\left(\sum_{j\ge 0} t_2^{jk}\right)\cr\cr
&=& \sum_{k\ge 1}\frac{u^k}{k}\,\frac{1}{(1-t_1^k)(1-t_2^k)}.
\eea
Differentiate with respect to $u$ to find
\bea
F'(u)&=&\sum_{k\ge 1} u^{k-1}\,\frac{1}{(1-t_1^k)(1-t_2^k)}F(u)\cr\cr
&=&\sum_{N\ge 0}\sum_{k\ge 1} u^{N+k-1}\,\frac{1}{(1-t_1^k)(1-t_2^k)}\tilde{Z}_N(t_1,t_2)
\eea
Directly from the definition of $F(u)$ we have
\bea
F'(u)&=&\sum_{N\ge 1} N \tilde{Z}_N(t_1,t_2) \,u^{N-1}.
\eea
Equating the coefficient of $u^{N-1}$ in these last two expressions yields the recursion relation
\bea
N_N(t_1,t_2) &=& \frac{1}{N}\sum_{m=1}^N 
\frac{\prod_{i=1}^N (1-t_1^i)(1-t_2^i)}{\prod_{i=1}^{N-m} (1-t_1^i)(1-t_2^i)}
\,\frac{N_{N-m}(t_1,t_2)}{(1-t_1^m)(1-t_2^m)}\,,
\eea
with the initial condition $N_0(t_1,t_2)=1$. With this formula it is straightforward to compute the Molien-Weyl partition function explicitly. We have also tested that the generating function
\bea
\prod_{i,j\ge 0}\frac{1}{1-t_1^{i_1} t_2^{i_2}\cdots t_d^{i_d} u}
&=& \sum_{N\ge 0} u^N \,\tilde{Z}_N(t_1,t_2,\cdots,t_d)\cr\cr
&=& \sum_{N\ge 0} u^N \,\frac{N_N(t_1,t_2,\cdots,t_d)}{\prod_{i=1}^N (1-t_1^i)(1-t_2^i)\cdots (1-t_d^i)}\,,
\eea
reproduces the $d>2$ Molien-Weyl functions. From this generating function we derive the following recursion relation
\bea
N_N(\{t_i\}) = \frac{1}{N}\sum_{m=1}^N 
\frac{\prod_{i=1}^N (1-t_1^i)(1-t_2^i)\cdots(1-t_d^i)}{\prod_{i=1}^{N-m} (1-t_1^i)(1-t_2^i)\cdots (1-t_d^i)}
\,\frac{N_{N-m}(\{t_i\})}{(1-t_1^m)(1-t_2^m)\cdots(1-t_d^m)}\cr
\eea
where $N_N(\{t_i\})=N_N(t_1,t_2,\cdots,t_d)$ and again $N_0(t_1,t_2,\cdots,t_d)=1$. By explicit computation with these formulas we find that the number of secondary invariants is given by $(N!)^{d-1}$.

\subsection{Palindromicity}\label{mwproperties}

The numerator of the Mollien-Weyl partition function is often palindromic. This property is useful in the construction of secondary invariants, as it provides a definite upper bound for the degree of secondary invariants. When applying our construction algorithm we do not need to search above this upper bound. In the case of multi-matrix models, starting from the contour integral for the Mollien-Weyl function \cite{Teranishi} proved that, for $U(2)$ invariants of 2$\times$2 matrices, the numerator of the Hilbert series is palindromic. Palindromicity was also observed  for finite $N$ partition functions of matrix models, connected to matrix quantum mechanics limits of ${\cal N}=4$ super Yang-Mills theory \cite{Harmark:2014mpa}.  See also \cite{Kristensson:2020nly} for further discussions of Hilbert series for matrix models. In Appendix \ref{MW} we prove that the partition function is palindromic for even $d$. This implies that, for even $d$ we have\footnote{We use $\tilde{Z}$ to denote the partition function with ground state energy set to zero and $Z$ to denote the partition function with ground state energy ${\omega\over 2}$ for each oscillator of frequency $\omega$. $\tilde{Z}$ that is directly related to the Hilbert series, while $Z$ is palindromic. The relation between the two is $Z(x)=x^{dN\over 2}\tilde{Z}(x)$ and $Z(t_1,t_2,\cdots,t_{2k})=t_1^{N\over 2}\cdots t_{2k}^{N\over 2}\tilde{Z}\left(\frac{1}{t_1},\frac{1}{t_2},\cdots,\frac{1}{t_{2k}}\right)$.}
\bea
Z(x) &=& Z\left(\frac{1}{x}\right)\label{palindromicformula}
\eea
where $x=e^{-\beta}$ with $\beta$ the inverse temperature. We can also consider a finer grading by introducing $\mu_a$ the chemical potential for quanta of the $a=1,2,...,d$ oscillators\footnote{Since we consider free oscillators in $d$ dimensions the quanta associated to the oscillator $a^{a\dagger}$ are conserved for each $a=1,2,\cdots,d$.}. In this case we obtain a more refined version of palindromicity, which reads
\bea
Z(t_1,\cdots,t_{2k})&=&Z\left(\frac{1}{t_1},\cdots,\frac{1}{t_{2k}}\right)
\eea
The physical interpretation of (\ref{palindromicformula}) is that it is related to temperature inversion symmetry $T\to -T$ and it has been observed in a number of other theories \cite{McGady:2017rzv,McGady:2018rmo}. For odd $d$ this is not a symmetry of the theory.

The partition function for $N$ bosons in $d$ dimensions takes the form
\bea
Z(x)&=&(x)^{dN\over 2}\,\,\frac{1+\sum_{n>0}^{n_{\rm max}} c_{n}x^n}{\prod_{m=1}^N(1-x^m)^d}
\eea
When $d$ is even, palindromicity implies the following equation for coefficients $c_{n,m}$
\bea
c_{n}&=&c_{n_{\rm max}-n}
\eea
and it gives the following equation for the biggest degree $n_{max}$
\bea
n_{max}=\frac{dN(N-1)}{2}
\eea
Thus we only need to search for secondary invariants of this degree or lower. Using the more refined version of palindromicity, it is clear that the largest multi degree is  $\frac{N(N-1)}{2}$ in each variable.

For further details and explicit results, the reader should consult Appendix \ref{MW}.

\section{Algorithm for the secondary invariants of $\mathbb{C}[V^{Nd}]^{S_N}$}\label{algorithm}

In this Section we present a new algorithm for the computation of secondary invariants, assuming that primary invariants are given. An algorithm to construct the secondary invariants, given the primary invariants has been given in \cite{king}. That algorithm uses the full machinery of Gr\"obner bases. A clear advantage of our algorithm is that it only uses elementary linear algebra.

Given a complete set of primary invariants (homogenous system of parameters), we first quotient out all polynomials in the primaries and then generate candidate invariants in increasing (multi)degree from the remaining trace generators. Linear relations among these candidates (trace identities) are extracted by an evaluation–interpolation step: choose generic points $p_k\in\mathbb{R}^{Nd}$ (or $\mathbb{C}^{Nd}$), form the matrix $M$ with entries $M_{k\ell}=I_\ell(p_k)$ where each column $I_\ell$ corresponds to a candidate invariant, and take as many points as needed so that $M$ is square (or tall). A full column rank indicates no relations in that degree; a nontrivial right-nullspace furnishes a basis of independent trace relations. Removing the dependent columns yields a linearly independent set of secondary invariants. The procedure proceeds degree by degree and terminates once the graded multiplicities match those predicted by the Molien–Weyl (Hilbert) series. As a consistency check, at any fixed multidegree we reconstruct all invariants from the obtained primaries and secondaries and verify -- again by rank over generic evaluations -- that no further relations occur. The following subsection provides a detailed, concrete case study.

The constraints summarized in (\ref{PermCayleyHamilton}) are the constrints produced by our evaluation-interpolation algorithm. For example, at $N=2$ it is easy to verify that (\ref{PermCayleyHamilton}) produces a degree (2,1) constraint given by
\bea
2(2,1)_I-2(1,1)_I(1,0_I)-(2,0)_I(0,1)_I+(1,0)_I^2(0,1)_I&=&0
\eea
It is simple to verify that this constraint is the null state of the matrix
\bea
\left[
\begin{matrix}
(2,1)_I(p_1) &(1,1)_I(1,0)_I(p_1) &(2,0)_I(0,1)_I(p_1) &(1,0)_I^2(0,1)_I(p_1)\\
(2,1)_I(p_2) &(1,1)_I(1,0)_I(p_2) &(2,0)_I(0,1)_I(p_2) &(1,0)_I^2(0,1)_I(p_2)\\
(2,1)_I(p_3) &(1,1)_I(1,0)_I(p_3) &(2,0)_I(0,1)_I(p_3) &(1,0)_I^2(0,1)_I(p_3)\\
(2,1)_I(p_4) &(1,1)_I(1,0)_I(p_4) &(2,0)_I(0,1)_I(p_4) &(1,0)_I^2(0,1)_I(p_4)
\end{matrix}
\right]
\eea
In more complicated examples, the constraints needed to eliminate redundant invariants are produced by carefully chosen linear combinations of the constraints following from (\ref{PermCayleyHamilton}). The evaluation-interpolation algorithm constructs the relevant constraint automatically, simplifying the analysis.

\subsection{Numerical Details}\label{Numerics}

In this section we consider the case $N=3$ and $d=2$, so that our invariants are functions of two $3\times 3$ matrices, $X^1$ and $X^2$. The method generalizes straightforwardly to other values of $N$ and $d$. To ease the notation from now on we simply indicate invariants by their multidegree, so that the invariant $(a,b)_I$ is given by ${\rm Tr}((X^1)^a(X^2)^b)$.

\noindent
{\bf Primary invariants:} The complete set of primary invariants is
\bea
\{(1,0)_I,\,(0,1)_I,\,(2,0)_I,\,(0,2)_I,\,(3,0)_I,\,(0,3)_I\}\,.
\eea

\noindent
{\bf Fundamental invariants beyond primaries:} After removing all primaries and their polynomial products, the remaining fundamental invariants are
\bea
\{(1,1)_I,\,(2,1)_I,\,(1,2)_I\}\,.
\eea
Candidate secondary invariants are constructed from the above set.

\noindent
{\bf Algorithmic strategy:}
\begin{itemize}
\item The algorithm proceeds degree by degree. At each degree, scan all possible multidegrees $(n,m)$ with $1\leq n,m\leq \tfrac{N(N-1)}{2}=3$.  
\item To test independence, invariants are evaluated numerically by assigning values to the six variables $\{x^1_i, x^2_i\}_{i=1}^3$, of the diagonal matrices $X^1$ and $X^2$.  
\item At a given multidegree $(n,m)$, construct matrix $M(n,m)$ whose columns correspond to candidate invariants and products of lower-degree primaries/secondaries, while the rows correspond to different numerical evaluations.  
\item The null space of $M(n,m)$ reveals trace relations. A full-rank matrix indicates no relation, while a non-trivial null space signals linear dependence among the candidates.  
\end{itemize}

\noindent
{\bf Step-by-step construction:}
\paragraph{Degree 0.} The trivial secondary invariant is $1$.  

\paragraph{Degree 1.} All invariants are primary.  

\paragraph{Degree 2, multidegree $(1,1)$.} 
Candidates: $\{(1,1)_I,\,(1,0)_I(0,1)_I\}$.  $M(1,1)$ is full rank, i.e.\ no trace relations. 
Thus $(1,1)_I$ is a genuine secondary invariant.  

\paragraph{Degree 3, multidegrees $(2,1)$ and $(1,2)$.} At $(2,1)$ the candidate secondary invariant is $(2,1)_I$. The complete set of invariants is $\{(2,1)_I,\,(2,0)_I(0,1)_I,\,(1,0)_I^2(0,1)_1,$ $\,(1,1)_I(1,0)_I\}$. $M(2,1)$ is full rank, so $(2,1)_I$ is a new secondary.  

At $(1,2)$ the situation is symmetric, and $(1,2)_I$ is also a new secondary.  

\paragraph{Degree 4, multidegrees $(2,2)$, $(1,3)$ and $(3,1)$.} At multidegree $(2,2)$ the secondary candidates are: $(2,2)_I$ and $(1,1)_I^2$. There are nine possible invariants that can be constructed and $M(2,2)$ has rank 8, indicating one trace relation. The null state of $M(2,2)$ gives the trace relation
\bea
&&-6(2,2)_I+(2,0)_I(0,2)_I-(1,0)_I^2(0,2)_I-(2,0)_I(1,0)_I^2+(1,0)_I^2(0,1)_I^2\cr
&&-4(1,1)_I(1,0)_I(0,1)_I+2(1,1)_I^2+4(2,1)_I(0,1)_I+4(1,2)_I(1,0)_I=0\,.
\eea
This relation allows us to eliminate $(2,2)_I$, leaving $(1,1)_I^2$ as a new secondary invariant.  At $(3,1)$ and $(1,3)$, there are no potential secondary invariants.  

\paragraph{Degree 5, multidegrees $(3,2)$ and $(2,3)$.} All potential secondary invariants are removed by trace relations.  

\paragraph{Degree 6, multidegree $(3,3)$.} There are two candidate secondary invariants given by: $(2,1)_I(1,2)_I$, and $(1,1)_I^3$.  There is a single relation, leaving $(1,1)_I^3$ as the only new secondary.  

\noindent
{\bf Final result:} For $N=3$, $d=2$, the complete set of secondary invariants is $\{(1,1)_I,\,(2,1)_I,\,(1,2)_I,\,(1,1)_I^2,\,(1,1)_I^3\}$.

At each step the construction matrices $M(n,m)$ and their null spaces identify trace relations, ensuring that only independent secondary invariants are retained. Whenever possible, we choose to keep multi-particle (reducible) secondary invariants. The result is not unique. For example, at degree 6 we could equally well have chosen to keep $(2,1)_I(1,2)_I$ as the secondary invariant, while discarding $(1,1)_I^3$.

There are some compelling tests that can be performed to confirm the above system of invariants: we can test that they generate a free module. Concretely, given the primary ($P_i$) and secondary ($S_j$) invariants we can construct the set of all operators of the form
\bea
\label{eq:prodofprimandonesec}
\prod_{i=1}^{dN} (P_i)^{n_i}S_j
\eea
that have a fixed multidegree. For a valid system of invariants, there are no relations between the operators in the set. We have tested this extensively for the invariant systems we have computed. As an example, using the system of invariants constructed above we find there are a total of 350 operators at multidegree $(8,8)$ and these are all independent -- a highly non-trivial test of the invariant system derived above. For further discussion of the tests we performed the reader is referred to Appendix \ref{numtests}.

\section{Invariants for $N$ and $d=2$}\label{dis2story}

In this Section we consider the construction of secondary invariants for the family of models with $d=2$ and arbitrary $N$. As we will see, this provides a concrete realization of finite $N$ cut offs and $q$-reducibility.

From the form of the Molien-Weyl functions (see Appendix \ref{MWPFs}) we learn that the invariants enjoy some non-trivial properties. The numerator of the Molien-Weyl partition function itself is the graded Hilbert series for the secondary invariants and it indicates a number of interesting properties enjoyed by the secondary invariants:
\begin{itemize}
\item {\bf Stability:} The complete set of invariants of length $\leq N$ is included among the primary and secondary invariants. In particular, the complete set of irreducible secondary invariants is the set of single--trace invariants with $\leq N$ matrices that are not primary invariants. The secondary invariants exhibit stability: as $N$ increases, the invariants constructed from at most $N$ matrices are unchanged.  
\item {\bf Palindromicity:} We have proved that the Molien--Weyl partition function is palindromic in Appendix \ref{MW}. Consequently, the number of invariants of multidegree $(n,m)$ equals the number of invariants of multidegree $\big(\tfrac{N(N-1)}{2}-n,\tfrac{N(N-1)}{2}-m\big)$. 
\item {\bf Total number:} The total number of secondary invariants is $N!$. 
\end{itemize}

From explicit examples, it soon becomes clear that the complete set of secondary invariants collapse into towers of the form:
\bea
({\rm base})_{(a,b)}(1,1)_I^k\qquad\qquad k=0,1,2,\cdots,{N(N-1)\over 2}-a-b
\eea
Here $({\rm base})_{(a,b)}$ is a secondary invariant, not necessarily irreducible, of bidegree $(a,b)$. It is possible to construct the Hilbert series for the $({\rm base})_{(a,b)}$s themselves. Using mathematica, we can construct graded Hilbert series of the form
\bea
g_N(t_1,t_2)&=&\sum_{a,b=1}^{a+b={N(N+1)\over 2}}c_{a,b}t_1^at_2^bx^{a+b}
\eea
which count the $({\rm base})_{(a,b)}$s by bidegree $(a,b)$ and degree $a+b$. These Hilbert series suggest the organization of the secondary invariants into towers retains much of the interesting structure we identified above. In particular, we have: 
\begin{itemize}
\item[(i)] {\bf Stability:} the graded counting of bases exhibits a similar stability as for the counting of invariants and the counting of secondary operators: the counting of bases of total degree $\le N$ agrees with the counting at $N=\infty$. 
\item[(ii)] {\bf Palindromic:} The graded polynomial multiplying each power of $x$ is palindromic. This is {\it not} to be confused with the palindromic property of the complete Molien-Weyl function. The palindromicity of the Molien-Weyl function related the coefficients of the polynomial that count different degrees. The palindromicity we have here is all at a single total degree.
\end{itemize}

At any value of $m<N$ we have the single trace operators with degree $(a,m-a)$ for $a=1,2,\cdots,m-1$. These are the building blocks of our construction - it is from these single trace operators that we construct everything else. To move from simply counting invariants to describing the specific operators we consider, introduce the polynomial
\bea
h_k(t_1,t_2)&=&\sum_{l=1}^{k-1}t_1^l t_2^{k-l}\qquad\qquad k\ge 3
\eea
Each term in this polynomials is a single trace operator of the advertised bidegree. Since the single trace operators are the building blocks for constructing the bases of the towers, these polynomials are the building blocks for the functions $g_N(x,t_1,t_2)$ introduced above. When we multiply two polynomials, we are constructing a polynomial that counts double trace operators. It is easy to see that when we multiply polynomials $h_{k_1}(t_1,t_2)h_{k_2}(t_1,t_2)\cdots h_{k_l}(t_1,t_2)$ with the labels $k_1,k_2,\cdots,k_l$ all distinct, the usual multiplication between polynomials correctly gives the counting of multitrace operators. This multiplication rule must, however, be modified when labels are repeated. As an example
\bea
h_3(t_1,t_2)=t_1t_2^2+t_1^2t_2\quad\leftrightarrow\qquad \{(1,2)_I,(2,1)_I\}
\eea
Using the usual product between polynomials we have
\bea
\left(h_3(t_1,t_2)\right)^2=t_1^2 t_2^4+2 t_1^3t_2^3+t_1^4t_2^2
\eea
which does not count the set of double trace operators that is given by
\bea
\{(1,2)_I^2,(1,2)_I(2,1)_I,(2,1)_I^2\}
\eea
The discrepancy arises because the usual multiplication between polynomials counts $(1,2)_I(2,1)_I$ and $(2,1)_I(1,2)_I$ (both of which appear if we simply multiply all elements of the set $\{(1,2)_I,(2,1)_I\}$ with itself) as distinct. We need to take a product that correctly drops duplicates. The product which deletes duplicates is easily implemented in mathematica. We denote this product by $*_d$.

In terms of these building blocks we can immediately write a formula for the stable part of $g_N(x,t_1,t_2)$ as follows
$$ $$
\bea
g_N^{\rm stable}(x,t_1,t_2)&=&1+\prod_{\substack{k_1,k_2,\cdots k_l\\ k_1+k_2+\cdots +k_l\le N}}h_{k_1}(t_1,t_2)*_d h_{k_2}(t_1,t_2)*_d \cdots *_d h_{k_l}(t_1,t_2)\, x^{k_1+k_2+\cdots +k_l}\cr
&&
\label{eq:gNstable}
\eea
To derive this formula, simply count all invariants constructed using fewer than $N$ $X^1$s and $X^2$s. Since these invariants are all composed from fewer than $N$ fields, no trace relations are possible and this matches the $N=\infty$ theory i.e. by definition this is the stable contribution. We stress that the above equation is to be understood as a specification of the operators that correspond to the bases of the stable secondary invariants: each term in each $h_{k}(t_1,t_2)$ corresponds to a specific single trace operator. The description in terms of $h_{k}(t_1,t_2)$ polynomials has an immediate translation into concrete operators.

The coefficient of $x^{N+1}$, which is the first non-stable contribution, also has a simple structure. The corresponding term is given by
\bea
\label{eq:gNn}
\prod_{\substack{k_1,k_2,\cdots,k_l\\ k_1+k_2+\cdots +k_l= N+1\\
l>1}}h_{k_1}(t_1,t_2)*_d h_{k_2}(t_1,t_2)*_d \cdots *_d h_{k_l}(t_1,t_2)\, x^{N+1}
\eea
The coefficient of $x^{N+2}$, the second non-stable contribution can be written as
\bea
&&\prod_{\substack{k_1,k_2,\cdots,k_l\\ k_1+k_2+\cdots +k_l= N+2\\
l>1}}h_{k_1}(t_1,t_2)*_d h_{k_2}(t_1,t_2)*_d \cdots *_d h_{k_l}(t_1,t_2)\, x^{N+2}
-\delta_{N,4}h_3*_d h_3x^{N+2}\cr\cr
\label{eq:gNnn}
&&\qquad\qquad\qquad\qquad-\delta_{N>4}(h_3*_d h_{N-1}|_{t_1^{N-2}t_2}+h_{N-2}*_d h_4|_{t_1t_2^3})x^{N+2}
\eea
The notation above deserves some explanation. $h_{N-1}|_{t_1^{N-2}t_2}$ stands for the term $t_1^{N-2}t_2$ from $h_{N-1}$, which corresponds to the invariant $(N-2,1)_I$. Thus, $h_3*_d h_{N-1}|_{t_1^{N-2}t_2}$ is a pair of double trace operators given by $\{(N-2,1)_I(1,2)_I,(N-2,1)_I(2,1)_I\}$. It is easy to verify that the terms we subtract appear in the leading product term, so that the above formula states exactly which operators are used as bases of towers at degree $N+2$. All subtractions above are double trace operators. In a similar way, the coefficient of $x^{N+3}$ is given by
\bea
&&\prod_{\substack{k_1,k_2,\cdots,k_l\\ k_1+k_2+\cdots +k_l= N+3\\
l>1}}h_{k_1}(t_1,t_2)*_d h_{k_2}(t_1,t_2)*_d \cdots *_d h_{k_l}(t_1,t_2)\, x^{N+3}-h_3*_d h_Nx^{N+3}\cr\cr
\label{eq:gNnnn}
&&\qquad\qquad\qquad-\delta_{N,5}h_4|_{t_1^3t_2}*_dh_4|_{t_1t_2^3}x^{N+3}
-h_{N-3}*_d h_3|_{t_1^1t_2^2}*_d h_3|_{t_1^2t_2^1}x^{N+3}
\eea
We are subtracting both double trace and triple trace terms. In additions, all terms subtracted are distinct. Finally, the coefficient of $x^{N+4}$ is given by
\bea
&&\delta_{N>4}\left(\prod_{\substack{k_1,k_2,\cdots,k_l\\ k_1+k_2+\cdots +k_l= N+4\\
l>1}}h_{k_1}(t_1,t_2)*_d h_{k_2}(t_1,t_2)*_d \cdots *_d h_{k_l}(t_1,t_2)\, x^{N+4}
-h_N*_d h_4 x^{N+4}\right)\cr\cr
\label{eq:gNnnnn}
&&\qquad\qquad\qquad-\delta_{N,6}h_4*_dh_3|_{t_1t_2^2}*_dh_3|_{t_1^2t_2}x^{N+4}
-\delta_{N>6}h_4|_{t_1^2t_2^2}*_d h_4*_dh_{N-4}x^{N+4}
\eea
While a systematic formula for the further corrections is not obvious, it is clear that there is no obstruction to writing a formula for the coefficient of higher powers $x^{N+k}$ $k>4$ that would be true at all $N$. This strongly suggests that there is no obstruction to writing down formulas, valid for any $N$, for the operators that give the secondary invariants at a given total degree. The above formulas were tested for $N=3,4,\cdots,13$ using mathematica.

Another instructive exercise is to follow the set of secondary invariants of a given total degree, as $N$ is varied. Two examples are given below.

\noindent
{\bf Evolution of the set of secondary invariants of degree $9$:}
\bea
N\le4 && 0\cr
&&\downarrow\cr
N=5 &&h_3*_dh_3*_dh_3\cr
&&\downarrow\cr
N=6 &&h_3*_dh_3*_dh_3+h_3*_dh_6\cr
&&\downarrow\cr
N=7 &&h_3*_dh_3*_dh_3+h_3*_dh_6+(h_5|_{t_1^3t_2^2}+h_5|_{t_1^2t_2^3})h_4\cr
&&\downarrow\cr
N=8 &&h_3*_dh_3*_dh_3+h_3*_dh_6+h_4*_dh_5\cr
&&\downarrow\cr
N\ge9 &&h_3*_dh_3*_dh_3+h_3*_dh_6+h_4*_dh_5+h_9
\eea
At $N=7$ the terms $h_5|_{t_1^3t_2^2}+h_5|_{t_1^2t_2^3}$ are the operators of these bidegrees from the set represented by $h_5$, i.e. $(3,2)_I$ and $(2,3)_I$. 

\noindent
{\bf Evolution of the set of secondary invariants of degree $10$:}
\bea
N\le 4 &&0\cr
&&\downarrow\cr
N=5 &&h_3|_{t_1^2t_2}*_dh_3|_{t_1^2t_2}*_dh_4|_{t_1^1t_2^3}+h_3|_{t_1t_2^2}*_dh_3|_{t_1t_2^2}*_dh_4|_{t_1^3t_2}\cr
&&\downarrow\cr
N=6 &&h_3*_dh_3*_dh_4+h_5*_dh_5\cr
&&\downarrow\cr
N=7 &&h_3*_dh_3*_dh_4+h_5*_dh_5+h_3*_dh_7\cr
&&\downarrow\cr
N=8 &&h_3*_dh_3*_dh_4+h_5*_dh_5+h_3*_dh_7+h_6*_dh_4|_{t_1^2t_2^2}+h_4*_dh_6|_{t_1^3t_2^3}\cr
&&\downarrow\cr
N=9 &&h_3*_dh_3*_dh_4+h_5*_dh_5+h_3*_dh_7+h_4*_dh_6\cr
&&\downarrow\cr
N\ge10 &&h_3*_dh_3*_dh_4+h_5*_dh_5+h_3*_dh_7+h_4*_dh_6+h_{10}
\eea
At $N=5$ we selected the operators $(2,1)_I(2,1)_I(1,3)_I$ and $(1,2)_I(1,2)_I(3,1)_I$ from the set given by $h_3*_dh_3*_dh_4$. This choice is not unique and we could have selected any two operators of bidegree $(5,5)$. At $N=8$, $h_4|_{t_1^2t_2^2}$ stands for the operator $(2,2)_I$ while $h_6|_{t_1^3t_2^3}$ stands for the operator $(3,3)_I$ and these choices are unique. 

The above functions at each $N$ show exactly what operator provide bases at each value of $N$. As $N$ increases more and more operators are included until, when we reach the stable limit, no new operators are included.

\section{Reducible and irreducible secondary invariants}\label{coinvariantalgebra}

In this Section we will give another perspective on our construction algorithm, by introducing the coinvariant algebra. This will give a deeper understanding of why the secondary invariants can always be chosen so that they are \emph{monomials} (power products) in a smaller set of ``irreducible'' secondary invariants.

We consider algebras $A$ of invariants, generated by a system of \emph{primary} and \emph{secondary} invariants. The primary invariants define a homogeneous system of parameters\footnote{A homogeneous system of parameters (hsop) in a graded algebra $A$ is a collection of algebraically independent homogeneous elements $P_1,\cdots ,P_r$ such that the Krull dimension of $A$ equals $r$.}. The homogeneous system of parameters $P_1,\dots,P_r$ generates a polynomial subring $A_P=\mathbb{C}[P_1,\dots,P_r]\subset A$, and (since $A$ is Cohen--Macaulay) the invariant ring $A$ is a free $A_P$--module. For our application we are working over the field of complex numbers $\mathbb{C}$. The \emph{Hironaka decomposition} implies the direct sum
\begin{equation}
  A \;=\; \bigoplus_{j=1}^m A_P \cdot S_j,\label{hironaka}
\end{equation}
where the $S_j$ are the \emph{secondary invariants}. 

\noindent
{\bf The coinvariant algebra:} Let $\mathfrak f=(P_1,\dots,P_r)$ be the ideal generated by the primaries, and consider the \emph{coinvariant algebra}\footnote{If $A$ is a ring and $I \subseteq A$ is an ideal, then $A/I$ denotes the \emph{quotient ring}: two elements of $A$ are identified if their difference lies in $I$. Concretely,
\[
\mathfrak f A = \Big\{\, \sum_i h_i P_i \;\Big|\; h_i \in A \,\Big\}.
\]
Thus the notation $A/\mathfrak f A$ means: take the ring $A$ and declare two elements of $A$ to be the same if they differ by something divisible by at least one of the $P_i$. Intuitively we are ``setting all the primaries equal to zero.''}\cite{Sturmfels}
\begin{equation}
  A_{\rm coinv} \;=\; A / \mathfrak f A.
\end{equation}
Modulo $\mathfrak f$, the coefficients from $A_P$ vanish, so the images $\overline{S}_j$ of the secondary invariants form a $\mathbb{C}$-basis\footnote{Every time we say ``a $\mathbb{C}$-basis,” we mean ``a basis as a vector space over the ground field $\mathbb{C}$.”} of $A_{\rm coinv}$. Thus, understanding secondary invariants is equivalent to understanding the structure of $A_{\rm coinv}$ as a graded $\mathbb{C}$--algebra.

If $A$ is not free over $A_P$, this statement can fail. To see this, suppose $A$ is a graded invariant ring and $A_P=\mathbb{C}[P_1,\dots,P_r]$ is the subring generated by a homogeneous system of parameters.

\emph{Case 1: $A$ Cohen--Macaulay (free over $A_P$).}  
Since $A$ admits a Hironaka decomposition (\ref{hironaka}), reducing modulo $\mathfrak f=(P_1,\dots,P_r)$ gives
\begin{equation}
A/\mathfrak f A \;\cong\; \bigoplus_{j=1}^m (A_P/\mathfrak f)\cdot \overline{S}_j
\;\cong\; \bigoplus_{j=1}^m \mathbb{C} \cdot \overline{S}_j.
\end{equation}
Thus the images $\overline{S}_j$ are linearly independent and form a $\mathbb{C}$--basis of the coinvariant algebra.

\emph{Case 2: $A$ is not Cohen--Macaulay (not free over $A_P$).}  Even though an hsop $\{P_1,\dots,P_r\}$ exists, $A$ need not be a free $A_P$--module.  In this case one may try to write $A$ as $\sum_j A_P \cdot S_j$, but the sum is not direct.  Linear relations among the putative secondaries can survive modulo $\mathfrak f$, so the images $\overline{S}_j$ may be linearly dependent or fail to span.  Hence they need not form a basis of $A/\mathfrak f A$.

\medskip
So, the property that the images of the secondary invariants form a $\mathbb{C}$--basis of the 
coinvariant algebra is equivalent to $A$ being Cohen--Macaulay (i.e.\ free over $A_P$).  
For invariant rings of reductive groups this always holds, by the Hochster--Roberts theorem~\cite{hrtheorem}.

Inside $A_{\rm coinv}$, call a homogeneous element \emph{irreducible} if it cannot be written as a product of two elements of positive degree. Choose a minimal homogeneous generating set $\{\overline{s}_1,\dots,\overline{s}_t\}$ for $A_{\rm coinv}$ as a $\mathbb{C}$--algebra, consisting of such irreducible elements.
Lifting each $\overline{s}_i$ to a representative $s_i\in A$ gives a set of \emph{irreducible secondary invariants}.

Now consider the surjective graded $C$–algebra homomorphism
\bea
\varphi:\ C[Y_1,\dots,Y_t]\twoheadrightarrow A,\qquad Y_i\mapsto s_i.
\eea
Let $\pi:A\twoheadrightarrow A_{\mathrm{coinv}}:=A/fA$ be the natural quotient map and set
\bea
\bar\varphi := \pi\circ\varphi:\ C[Y_1,\dots,Y_t]\twoheadrightarrow A_{\mathrm{coinv}}.
\eea
Let $J=\ker(\bar\varphi)$. By Gröbner basis theory, the set of standard monomials in
$C[Y]/J$ (i.e. monomials not in the initial ideal $\mathrm{in}(J)$ for some term order)
maps to a $C$–basis of $A_{\mathrm{coinv}}$. Equivalently, the elements
\bea
s^\alpha := s_1^{\alpha_1}\cdots s_t^{\alpha_t},
\eea
form a $C$–basis of $A_{\mathrm{coinv}}$.

Finally, lift each $\overline{s}^\alpha$ to the corresponding power product $s^\alpha = s_1^{\alpha_1}\cdots s_t^{\alpha_t}\in A$. Because the images form a basis of $A_{\rm coinv}$, the set $\{s^\alpha\}$ is linearly independent modulo $\mathfrak f A$ and spans $A$ as an $A_P$--module. Therefore $\{s^\alpha\}$ is a valid $A_P$--basis of $A$, i.e.\ a set of secondary invariants.

We conclude that:
\begin{quote}
  \emph{Secondary invariants can always be chosen so that each is a power product of irreducible secondary invariants. This is true for any algebra of invariants that admits a Hironaka decomposition.}
\end{quote}
This follows from the fact that the coinvariant algebra $A_{\rm coinv}$ is finitely generated by irreducible elements, and that a $\mathbb{C}$--basis is given by standard monomials in them. Lifting these monomials back to $A$ yields the desired Hironaka decomposition.

\section{Conclusions}\label{discussion}

We have presented a purely evaluation--interpolation algorithm that constructs a free module basis of secondary invariants degree by degree. The algorithm forms numerical matrices whose nullspaces reproduce the trace relations. The algorithm terminates when the graded multiplicities of the constructed secondary invariants match the Molien–Weyl (Hilbert) series. This reduces the problem to elementary linear algebra and avoids Gr\"obner basis machinery. 

In the $S_N$ model of $N$ bosons in $d$ spatial dimensions--where the relevant matrices commute--we have implemented the algorithm explicitly. For $N=3,d=2$ we reproduce the expected primaries and isolated the genuine secondaries at each bidegree. As a stringent check of our results, wehave verfied that all 350 operators at multidegree $(8,8)$ are independent. 

We have also derived practical recursions for the Molien–Weyl numerators in any $d$, and established that the number of secondary invariants is $(N!)^{d-1}$. These inputs provide precise end-conditions for the construction algorithm. We have also proved that for even $d$ the partition function is palindromic, which implies a sharp upper bound on secondary degrees and a pairing of multiplicities across bidegrees. Together with the finite-length generation by single traces, this yields a clear notion of “stable” sectors that persist with increasing $N$. 

Our results prove that the set of secondary invariants organizes into towers built by repeatedly multiplying a base secondary by the $(1,1)_I$ generator up to a finite, $N$-dependent ceiling. This explicitly realizes $q$-reducibility: beyond degree-dependent cutoffs, products that would naively generate new states become reducible over the primaries and existing secondary invariants. In the emergent-Fock-space picture, primary invariants furnish free modes while secondary invariants act linearly. $q$-reducibility implements a finite-$N$ occupation-number cutoff, discarding multiparticle states as a consequence of finite-$N$ gravitational redundancy. Collective field theory provides a dynamical setting in which these constraints commute with the Hamiltonian and can be imposed as operator equations. 

The construction algorithm we have presented can also be applied to matrix model quantum mechanics. However, in that setting one would need to determine both the primary and secondary invariants using the algorithm. Although this is more involved, they key idea of generating trace relation using a purely evaluation--interpolation algorithm is applicable. The problem again reduces to simple linear algebra, but due to the massive number of primary and secondary invariants, it is most likely only possible to determine the complete system of invariants for small values of $N$.

Finally, we have pointed out that the coinvariant-algebra perspective may be useful to explore the structure of the space of secondary invariants.  This may well be a fruitful avenue for future study given that the secondary invariants are related to non-perturbative states of the theory,
 
\begin{center} 
{\bf Acknowledgements}
\end{center}
%We would like to thank ... for discussions. 
The work of RdMK is supported by a start up research fund of Huzhou University, a Zhejiang Province talent award and by a Changjiang Scholar award. AR acknowledges financial support from: the Department of Science, Technology, and Innovation (DSTI), South Africa, through COST Action CA22113, Fundamental Challenges in Theoretical Physics; the DSI/NRF South African Research Chairs Initiative (SARChI) Research Chair in Theoretical Particle Cosmology (Grant No. 78554); MITP bursary and extends sincere thanks to the Supergravity group in the Department of Applied Science and Technology (DISAT), Politecnico di Torino—led by Prof. Mario Trigiante—for their hospitality and partial support during this work.

\begin{appendix}

\section{Comments on the Molien-Weyl partition function}\label{MW}

In this section, we examine the Molien–Weyl partition function for a system of $N$ bosons in $d$ dimensions, which coincides with the partition function of the free theory with a harmonic oscillator external potential. The Molien–Weyl partition function is palindromic when $d$ is even, but not when $d$ is odd. Palindromicity of the partition function is closely tied to temperature inversion symmetry. In particular, \cite{McGady:2017rzv,McGady:2018rmo} present arguments suggesting that finite-temperature path integrals of quantum field theories should in general remain invariant under the reflection $\beta\to-\beta$. One result of this Appendix shows that systems of $N$ bosons in odd dimensions furnish an infinite family of counterexamples -- quantum mechanical models whose partition functions explicitly break the $\beta\to-\beta$ symmetry.

\subsection{The defining (or natural) representation of $S_n$}

In this section, we derive a formula for the determinant of an arbitrary element in the natural representation of $S_n$. This is used to derive the central identity needed to prove that the Molien-Weyl function is palindromic for even $d$.

The defining representation (also called the natural representation) of $S_n$ is an $n$ dimensional reducible representation, given by the direct sum of the trivial and the standard representations. Denoting the natural representation of $S_n$ by ${\rm nat}$, we have
\bea
\Gamma_{\rm nat}&=&{\tiny \yng(6)}\,\,\,\,\oplus\,\,\,\,{\tiny\yng(5,1)}
\eea 
where there are $n$ boxes in the first Young diagram and $n-1$ boxes in the first row of the second Young diagram. The natural representation is defined as follows
\bea
\Gamma_{\rm nat}(\sigma): x^a_i&\to&x^a_{\sigma(i)}\qquad \sigma\in S_n
\eea
Thus, for example, we have
\bea
\Gamma_{\rm nat}\left((12)\right)&=&\left[\begin{array}{cccccc}
0 &1 &0 &0 &\cdots &0\\
1 &0 &0 &0 &\cdots &0\\
0 &0 &1 &0 &\cdots &0\\
0 &0 &0 &1 &\cdots &0\\
\vdots &\vdots &\vdots &\vdots &\ddots &\vdots\\
0 &0 &0 &0 &\cdots &1
\end{array}\right]
\eea
i.e. we simply swapped the first two rows of the $n\times n$ identity matrix $1_{n\times n}$. The determinant of the identity matrix is $1$. Since the determinant changes sign under swapping rows, we immediately learn that
\bea
\det \Big(\Gamma_{\rm nat}\left((12)\right)\Big)&=&-1.
\eea
For any two elements $\sigma_1$ and $\sigma_2$ in the same conjugacy class we know that we can write $\sigma_1=\rho\sigma_2\rho^{-1}$ for some $\rho\in S_n$. Thus, in our matrix representation we have
\bea
\Gamma_{\rm nat}(\sigma_1)&=&\Gamma_{\rm nat}(\rho)\Gamma_{\rm nat}(\sigma_2)\Gamma_{\rm nat}(\rho^{-1})\,\,=\,\,\Gamma_{\rm nat}(\rho)\Gamma_{\rm nat}(\sigma_2)\Gamma_{\rm nat}(\rho)^{-1}
\eea
which proves that all matrices representing group elements in the same conjugacy class have identical eigenvalues\footnote{This conclusion is true for any matrix representation, not just $\Gamma_{\rm nat}$.}. Thus, for any two cycle we have
\bea
\det \Big(\Gamma_{\rm nat}\left((\cdot\cdot)\right)\Big)&=&-1.
\eea
Using the following facts
\begin{itemize}
\item The determinant of a product of matrices is the product of the determinants of each matrix. 
\item Every element $\sigma\in S_n$ can be decomposed into a product of two cycles.
\item The parity $\pi(\sigma)$ of permutation $\sigma$ is defined as the number of two cycles in the decomposition, taken modulo 2. Although the expression for a general element in terms of two cycles is not unique, the different decompositions all have the same parity.
\end{itemize}
we easily deduce
\bea
\det \Big(\Gamma_{\rm nat}\left(\sigma\right)\Big)&=&(-1)^{\pi(\sigma)}
\eea
Note that $\pi(\sigma)=\pi(\sigma^{-1})$ and $\pi(\sigma)$ is either 0 or 1. Using the above formula, we easily prove the identity we will use below
\bea
\det\Big(t1_{n\times n}-\Gamma_{\rm nat}\left(\sigma\right)\Big)&=&(-1)^{\pi(\sigma)}\det (\Gamma_{\rm nat}\left(\sigma^{-1}\right))\det\Big(t1_{n\times n}-\Gamma_{\rm nat}\left(\sigma\right)\Big)\cr\cr
&=&(-1)^{\pi(\sigma)}\det\Big(t\Gamma_{\rm nat}\left(\sigma^{-1}\right)-1_{n\times n}\Big)\cr\cr
&=&(-1)^{\pi(\sigma)+n}\det\Big(1_{n\times n}-t\Gamma_{\rm nat}\left(\sigma^{-1}\right)\Big)\label{ImpIdent}
\eea

\subsection{Derivation of the Molien-Weyl formula}

The Molien-Weyl formula computes the oscillator partition function. In this section we will give a derivation of the Molien-Weyl formula using a straight forward evaluation of the partition function. It gives an alternative to the derivation usual presented in the invariant theory of finite groups -- see for instance \cite{Sturmfels}. The simplest way to construct the states of free oscillator is by using creation operators, $a_i^{a\dagger}$ $a=1,2,\cdots,d$, $i=1,2,\cdots,N$. as usual. These oscillators transform in the natural representation $\Gamma_{\rm nat}$ of $S_N$. Subtract off the ground state energy and use units in which the energy spacing of each oscillator is 1. The state formed by acting with $n_a$ $a_i^{a\dagger}$'s has energy $\sum_{a=1}^d n_a$. Thus the partition function is
\bea
\tilde{Z}(x)&=&\sum_{n_1=0}^\infty\sum_{n_2=0}^\infty\cdots\sum_{n_d=0}^\infty x^{n_1+n_2+\cdots+n_d}\#(n_1,n_2,\cdots,n_d)\label{pf1}
\eea
where $x=e^{-\beta}$ and $\#(n_1,n_2,\cdots,n_d)$ is the number of singlets that can be formed by acting with $n_a$ $a_i^{a\dagger}$'s. This number is most easily computed using characters. The representation produced by acting with $n$ $a_i^\dagger$'s is given by the symmetric product of $n$ copies of the natural representation ${\rm sym}^n_{\Gamma_{\rm nat}}$. Thus, the state created by acting with $n_a$ $a_i^{a\dagger}$'s belongs to the representation ${\rm sym}^{n_1}_{\Gamma_{\rm nat}}\otimes {\rm sym}^{n_2}_{\Gamma_{\rm nat}}\otimes\cdots\otimes{\rm sym}^{n_d}_{\Gamma_{\rm nat}}$. Thus, by character orthogonality we have
\bea
\#(n_1,n_2,\cdots,n_d)&=&{1\over n!}\sum_{\sigma\in S_n}\chi_{{\rm sym}^{n_1}_{\Gamma_{\rm nat}}\otimes {\rm sym}^{n_2}_{\Gamma_{\rm nat}}\otimes\cdots\otimes{\rm sym}^{n_d}_{\Gamma_{\rm nat}}}(\sigma)\chi_{\rm singlet}(\sigma^{-1})\label{singnum}
\eea
This can be evaluated using the characters $\chi_{\rm singlet}(\sigma^{-1})=1$ and the character for the symmetric product of $r$ copied of an arbitrary representation $R$
\bea
\chi_{{\rm sym}^n_R}(\sigma)&=&\mu \int d^{d_R}\bar{y}\int d^{d_R}\, e^{-\sum_{i=1}^{d_R}y_i\bar{y}_i}
\left(\sum_{j,k=1}^{d_R}y_j\Gamma_R(\sigma)_{jk}\bar{y}_k\right)^n
\eea
where $d_R$ is the dimension of representation $R$ and $\mu$ is fixed by
\bea
1&=&\mu \int d^{d_R}\bar{y}\int d^{d_R}\, e^{-\sum_{i=1}^{d_R}y_i\bar{y}_i}
\eea
Inserting (\ref{singnum}) into (\ref{pf1}) we can do the sum over the $n_a$, and then the integral over the $y_i,\bar{y}_i$, which produces a product of inverse determinants. The final result is
\bea
\tilde{Z}(x)&=&{1\over n!}\sum_{\sigma\in S_n}{1\over \det(1_{n\times n}-x\Gamma_{\rm nat}(\sigma))^d}
\eea
One can repeat the derivation by including a chemical potential $\mu_a$ for each $a_i^{a\dagger}$. The formula for this refined partition function 
\bea
\tilde{Z}_{\rm ref}(t_1,t_2,\cdots,t_d)&=&{1\over n!}\sum_{\sigma\in S_n}{1\over \prod_{a=1}^d\det(1_{n\times n}-t_a\Gamma_{\rm nat}(\sigma))}
\eea
is obtained exactly as above. If we reinstate the ground state energies, we have
\bea
Z(x)&=&{x^{Nd\over 2}\over n!}\sum_{\sigma\in S_n}{1\over \det(1_{n\times n}-x\Gamma_{\rm nat}(\sigma))^d}
\eea
and
\bea
Z_{\rm ref}(t_1,t_2,\cdots,t_d)&=&{(t_1t_2\cdots t_d)^{N\over 2}\over n!}\sum_{\sigma\in S_n}{1\over \prod_{a=1}^d\det(1_{n\times n}-t_a\Gamma_{\rm nat}(\sigma))}
\eea

\subsection{$T$-inversion symmetry}

Under $T$-inversion we have $x\to x^{-1}$. Consider the partition function evaluated at $x^{-1}$:
\bea
Z(x^{-1})&=&{x^{-Nd\over 2}\over n!}\sum_{\sigma\in S_n}{1\over \det(1_{n\times n}-x^{-1}\Gamma_{\rm nat}(\sigma))^d}\cr\cr
&=&{x^{-Nd\over 2}x^{Nd}\over n!}\sum_{\sigma\in S_n}{1\over \det(x1_{n\times n}-\Gamma_{\rm nat}(\sigma))^d}\cr\cr
&=&{x^{Nd\over 2}\over n!}(-1)^{Nd}\sum_{\sigma\in S_n}{(-1)^{d\pi(\sigma)}\over \det(1_{n\times n}-x\Gamma_{\rm nat}(\sigma))^d}
\eea
where we used (\ref{ImpIdent}) to get to the last line. If $d$ is even we know that $(-1)^{d\pi(\sigma)}=(-1)^{Nd}=1$
and
\bea
Z(x^{-1})&=&{x^{Nd\over 2}\over n!}\sum_{\sigma\in S_n}{1\over \det(1_{n\times n}-x\Gamma_{\rm nat}(\sigma))^d}
\,\,=\,\, Z(x)
\eea
Explicit computation confirms that for even $d$ $Z(x)$ is invariant $x\to x^{-1}$ and that the numerator of $\tilde{Z}(x)$ is palindromic. For odd $d$ $(-1)^{d\pi(\sigma)}\ne 1$ and explicit computation of the Molien-Weyl partition function shows $Z(x)$ is {\it not} invariant under $x\to x^{-1}$ and that the numerator of $\tilde{Z}(x)$ is not palindromic.

\subsection{Mollien-Weyl partition functions}\label{MWPFs}

\noindent
{\bf $N=3$ and $d=2$:}
\bea
\tilde{Z}(t_1,t_2)&=&\frac{1+t_1 t_2+t_1^2 t_2+t_1 t_2^2+t_1^2 t_2^2+t_1^3 t_2^3}{(1-t_1) (1-t_1^2) (1-t_1^3) (1-t_2) (1-t_2^2) (1-t_2^3)}\label{hilbertseries}
\eea
\bea
Z(t_1,t_2)&=&(t_1 t_2)^{3\over 2}\frac{1+t_1 t_2+t_1^2 t_2+t_1 t_2^2+t_1^2 t_2^2+t_1^3 t_2^3}{(1-t_1) (1-t_1^2) (1-t_1^3) (1-t_2) (1-t_2^2) (1-t_2^3)}
\eea

\noindent
{\bf $N=3=d$:}
\bea
\tilde{Z}(t_1,t_2,t_3)=\frac{N(t_1,t_2,t_3)}{(1-t_1)(1-t_1^2)(1-t_1^3)(1-t_2)(1-t_2^2)(1-t_2^3)(1-t_3)(1-t_3^2)(1-t_3^3)}
\eea
where
\bea
N(t_1,t_2,t_3)&=&1+t_1 t_2+t_1^2 t_2+t_1 t_2^2+t_1^2 t_2^2+t_1^3 t_2^3+t_1 t_3+t_1^2 t_3+t_2 t_3+t_1 t_2 t_3+t_1^2 t_2 t_3\cr\cr
&&+t_1^3 t_2 t_3+t_2^2 t_3+t_1 t_2^2 t_3+t_1^2 t_2^2 t_3+t_1^3 t_2^2 t_3+t_1 t_2^3 t_3+t_1^2 t_2^3 t_3+t_1 t_3^2+t_1^2 t_3^2\cr\cr
&&+t_2 t_3^2+t_1 t_2 t_3^2+t_1^2 t_2 t_3^2+t_1^3 t_2 t_3^2+t_2^2 t_3^2+t_1 t_2^2 t_3^2+t_1^2 t_2^2 t_3^2+t_1^3 t_2^2 t_3^2+t_1 t_2^3 t_3^2\cr\cr
&&+t_1^2 t_2^3 t_3^2+t_1^3 t_3^3+t_1 t_2 t_3^3+t_1^2 t_2 t_3^3+t_1 t_2^2 t_3^3+t_1^2 t_2^2 t_3^3+t_2^3 t_3^3
\eea
Note that the numerator is not palindromic.

\noindent
{\bf $N=4$ and $d=2$:}
\bea
\tilde{Z}(t_1,t_2)&=&\frac{N_4(t_1,t_2)}{(1-t_1)(1-t_1^2)(1-t_1^3)(1-t_1^4)(1-t_2)(1-t_2^2)(1-t_2^3)(1-t_2^4)}\label{Hd2N4}
\eea
where
\bea
N_4(t_1,t_2)&=&1 + t_1 t_2 + t_1^2 t_2 + t_1^3 t_2 + t_1 t_2^2 + 2 t_1^2 t_2^2 +
     t_1^3 t_2^2 + t_1^4 t_2^2 + t_1 t_2^3 + t_1^2 t_2^3 + 2 t_1^3 t_2^3\cr\cr 
     &&+ t_1^4 t_2^3 + t_1^5 t_2^3 + t_1^2 t_2^4 + t_1^3 t_2^4 + 2 t_1^4 t_2^4 + 
    t_1^5 t_2^4 + t_1^3 t_2^5 + t_1^4 t_2^5 + t_1^5 t_2^5 + 
    t_1^6 t_2^6
\eea

\bea
\tilde Z(t_1,t_2)&=&(t_1 t_2)^{4\over 2}\frac{N_4(t_1,t_2)}{(1-t_1)(1-t_1^2)(1-t_1^3)(1-t_1^4)(1-t_2)(1-t_2^2)(1-t_2^3)(1-t_2^4)}
\eea

\section{Numerical validation of the generating invariants}\label{numtests}

Invariants $I$ for the space of gauge invariant operators can all be written in the form
\bea
\label{eq:InvForm}
	I_{A} = P^{n_1}_{1}P^{n_2}_{2}\cdots P^{n_r}_{r}S_{\gamma},
\eea
where the label $A = \{ n_{1} , n_{2} , \cdots , n_{r} , \gamma \}$. In this section, we verify that all distinct operators of the form (\ref{eq:InvForm}) are linearly independent, i.e., there are no trace relations between these operators. Trace relations can only occur amongst operators at a given multi-degree. Thus, we collect all operators of the form (\ref{eq:InvForm}) of a specific multi-degree and test for trace relations. To do this, we build square matrices $M_{kA} = I_{A}(p_{k})$ from these invariants and verify that these matrices have full rank. A full rank indicates that there are no trace relations amongst the invariants and that they are independent. We perform our tests for $N=3,4,5$ and $d=2$. Lastly, all secondaries were obtained using equations (\ref{eq:gNstable}) - (\ref{eq:gNnnnn}) except for the $N=5$ degree 10 secondaries, where we used $\{(2,1)^{2}_{I}(1,3)_{I} , (1,2)^{2}_{I}(3,1)_{I} \}$. This is simply because, at $N=5$, degree 10 is the first non-stable contribution to $g_{N}(x,t_1,t_2)$ which goes beyond the formulas (\ref{eq:gNstable}) - (\ref{eq:gNnnnn}).

\subsection{$N=3$, $d=2$}

The primaries are given in (\ref{eq:Primaries}) and applying the algorithm described in section \ref{algorithm} yields the following secondary invariants
\bea
\label{eq:SecNeq3}
	\hbox{Secondaries } &=& \{ 1 , (1,1)_{I} , (2,1)_{I}  , (1,2)_{I}  ,  (1,1)^{2}_{I} , (1,1)^{3}_{I}   \}.
\eea
In Table \ref{TabNeq3checks} we present our results. We only need to consider all possible operators of the form (\ref{eq:InvForm}) having a specific bi-degree. At each bi-degree considered we give the total number of operators, which is also the size of the matrix $M_{kA}(p_{k})$, and the rank of $M_{kA}(p_{k})$. A full rank implies that there are no trace relations and that the operators at that bi-degree are all independent. 
\begin{table}[htp]
\caption{Table showing number of operators and rank of $M_{kA}$ for various bi-degrees at $N=3$. For each case, we obtain full rank indicating that all operators are independent.}
\begin{center}
\begin{tabular}{|c|c|c|c|}
\hline
$N$ & bi-degree & Number of operators & Rank \\
\hline
3 &(3,3)&19&19\\
\hline
 &(4,4)&42&42\\
\hline
 &(5,3)&38&38\\
\hline
 &(5,5)&78&78\\
\hline
 &(6,4)&76&76\\
\hline
 &(6,6)&139&139\\
\hline
 &(7,5)&132&132\\
\hline
 &(8,4)&120&120\\
\hline
 &(7,7)&224&224\\
\hline
 &(8,8)&350&350\\
\hline
\end{tabular}
\end{center}
\label{TabNeq3checks}
\end{table}%

\subsection{$N=4$, $d=2$}
We repeat the above procedure now for $N=4$ and $d=2$. The primaries are again given by (\ref{eq:Primaries}). The algorithm described in section \ref{algorithm} yields the following 24 secondary invariants
\bea
\label{eq:SecNeq4}
	\hbox{Secondaries } &=& \Big\{ 1 , (1,1)_{I} ,  (1,1)^{2}_{I} ,  (1,1)^{3}_{I} ,  (1,1)^{4}_{I} ,  (1,1)^{5}_{I}  ,  (1,1)^{6}_{I} \nonumber \\
					&& (1,2)_{I} , (1,1)_{I}(1,2)_{I} ,  (1,1)^{2}_{I}(1,2)_{I} ,  (1,1)^{3}_{I}(1,2)_{I} , \nonumber \\
					&& (2,1)_{I} , (1,1)_{I}(2,1)_{I} ,  (1,1)^{2}_{I}(2,1)_{I} ,  (1,1)^{3}_{I}(2,1)_{I} , \nonumber \\
					&& (1,3)_{I} , (1,1)_{I}(1,3)_{I},  (1,1)^{2}_{I}(1,3)_{I} , \nonumber\\
					&& (2,2)_{I} , (1,1)_{I}(2,2)_{I},  (1,1)^{2}_{I}(2,2)_{I} , \nonumber\\
					&& (3,1)_{I} , (1,1)_{I}(3,1)_{I},  (1,1)^{2}_{I}(3,1)_{I}  \Big\} 
\eea
In Table \ref{TabNeq4checks} we present our checks carried out for a variety of bi-degrees.
\begin{table}[htp]
\caption{Table showing number of operators and rank of $M_{kA}$ for various bi-degrees at $N=4$. For each case, we obtain full rank indicating that all operators are independent.}
\begin{center}
\begin{tabular}{|c|c|c|c|}
\hline
$N$ & bi-degree & Number of operators & Rank \\
\hline
4 &(4,4)&74&74\\
\hline
 &(5,5)&168&168\\
\hline
 &(6,6)&363&363\\
\hline
 &(7,5)&342&342\\
\hline
 &(7,6)&503&503\\
\hline
 &(7,7)&703&703\\
\hline
 &(8,6)&683&683\\
\hline
&(9,5)&606&606\\
\hline
 &(8,8)&1297&1297\\
\hline
\end{tabular}
\end{center}
\label{TabNeq4checks}
\end{table}%

\subsection{$N=5$, $d=2$ }
Lastly, we present our checks for $N=5$ and $d=2$. The primaries are again given by (\ref{eq:Primaries}). The algorithm yields the following secondary 120 invariants
\bea
\label{eq:SecNeq5}
	\hbox{Secondaries } &=& \Big\{ 1 , (1,1)_{I} ,  (1,1)^{2}_{I} , \cdots (1,1)^{10}_{I} , \nonumber \\
					&& (1,2)_{I} , (1,1)_{I}(1,2)_{I} ,  (1,1)^{2}_{I}(1,2)_{I} , \cdots   (1,1)^{7}_{I}(1,2)_{I} , \nonumber \\
					&& (2,1)_{I} , (1,1)_{I}(2,1)_{I} ,  (1,1)^{2}_{I}(2,1)_{I} , \cdots   (1,1)^{7}_{I}(2,1)_{I} , \nonumber \\
					&& (1,3)_{I} , (1,1)_{I}(1,3)_{I}, \cdots ,  (1,1)^{6}_{I}(1,3)_{I} , (2,2)_{I} , (1,1)_{I}(2,2)_{I}, \cdots , (1,1)^{6}_{I}(2,2)_{I} , \nonumber\\
					&& (3,1)_{I} , (1,1)_{I}(3,1)_{I},  \cdots , (1,1)^{6}_{I}(3,1)_{I} , (1,4)_{I} , (1,1)_{I}(1,4)_{I}, \cdots ,  (1,1)^{5}_{I}(1,4)_{I} , \nonumber\\
					&& (2,3)_{I} , (1,1)_{I}(2,3)_{I}, \cdots , (1,1)^{5}_{I}(2,3)_{I} , (3,2)_{I} , (1,1)_{I}(3,2)_{I},  \cdots , (1,1)^{5}_{I}(3,2)_{I} , \nonumber\\
					&& (4,1)_{I} , (1,1)_{I}(4,1)_{I},  \cdots , (1,1)^{5}_{I}(4,1)_{I} , (2,1)^{2}_{I} , (1,1)_{I}(2,1)^{2}_{I}, \cdots , (1,1)^{4}_{I}(2,1)^{2}_{I},\nonumber\\
					&& (2,1)_{I}(1,2)_{I} , (1,1)_{I}(2,1)_{I}(1,2)_{I}, \cdots , (1,1)^{4}_{I}(2,1)_{I}(1,2)_{I},\nonumber\\
					&& (1,2)^{2}_{I} , (1,1)_{I}(1,2)^{2}_{I}, \cdots , (1,1)^{4}_{I}(1,2)^{2}_{I},\nonumber\\
					&& (2,1)_{I}(2,2)_{I} , \cdots , (1,1)^{3}_{I}(2,1)_{I}(2,2)_{I} , (1,2)_{I}(2,2)_{I} , \cdots , (1,1)^{3}_{I}(1,2)_{I}(2,2)_{I},\nonumber\\
					&& (3,1)^{2}_{I}, (1,1)_{I}(3,1)^{2}_{I} , (1,1)^{2}_{I}(3,1)^{2}_{I} ,  (3,1)_{I}(2,2)_{I}, (1,1)_{I}(3,1)_{I}(2,2)_{I} ,\nonumber\\
					&&  (1,1)^{2}_{I}(3,1)_{I}(2,2)_{I}, (2,2)^{2}_{I}, (1,1)_{I}(2,2)^{2}_{I} , (1,1)^{2}_{I}(2,2)^{2}_{I} ,  (1,3)_{I}(2,2)_{I}, \nonumber\\
					&& (1,1)_{I}(1,3)_{I}(2,2)_{I} , (1,1)^{2}_{I}(1,3)_{I}(2,2)_{I} , (1,3)^{2}_{I}, (1,1)_{I}(1,3)^{2}_{I} , (1,1)^{2}_{I}(1,3)^{2}_{I} , \nonumber\\
					&& (2,1)^{3}_{I} , (1,1)_{I}(2,1)^{3}_{I}   ,  (2,1)^{2}_{I}(1,2)_{I} , (1,1)_{I}(2,1)^{2}_{I}(1,2)_{I} , \nonumber\\
					&& (1,2)^{3}_{I} , (1,1)_{I}(1,2)^{3}_{I}   ,  (1,2)^{2}_{I}(2,1)_{I} , (1,1)_{I}(1,2)^{2}_{I}(2,1)_{I} \nonumber\\
					&& (3,1)_{I}(1,2)^{2}_{I} , (2,1)^{2}_{I}(1,3)_{I}\Big\} . 
\eea
In Table \ref{TabNeq5checks} we present our checks carried out for a variety of bi-degrees.
\begin{table}[htp!]
\caption{Table showing number of operators and rank of $M_{kA}$ for various bi-degrees at $N=5$. For each case, we obtain full rank indicating that all operators are independent.}
\begin{center}
\begin{tabular}{|c|c|c|c|}
\hline
$N$ & bi-degree & Number of operators & Rank \\
\hline
 5&(5,5)&248&248\\
\hline
 &(6,6)&614&614\\
\hline
 &(7,5)&576&576\\
\hline
 &(8,4)&497&497\\
\hline
 &(7,7)&1367&1367\\
\hline
 &(8,6)&1319&1319\\
\hline
&(9,5)&1151&1151\\
\hline
\end{tabular}
\end{center}
\label{TabNeq5checks}
\end{table}%
It is particularly striking that, for example, at bi-degree (7,7) we constructed 1367 invariants of the form (\ref{eq:InvForm}) and verified that all of them were indeed linearly independent.

\end{appendix}


\begin{thebibliography}{60}

\bibitem{deMelloKoch:2025ngs} 
R.~de Mello Koch and A.~Jevicki, ``Structure of loop space at finite N,'' JHEP \textbf{06} (2025), 011
doi:10.1007/JHEP06(2025)011 [arXiv:2503.20097 [hep-th]].

\bibitem{deMelloKoch:2025qeq}
R.~de Mello Koch, M.~Kim and H.~J.~R.~Van Zyl, ``From Symmetry to Structure: Gauge-Invariant Operators in Multi-Matrix Quantum Mechanics,'' [arXiv:2507.01219 [hep-th]].

\bibitem{Sundborg:1999ue}
B.~Sundborg, ``The Hagedorn transition, deconfinement and N=4 SYM theory,'' Nucl. Phys. B \textbf{573} (2000), 349-363 doi:10.1016/S0550-3213(00)00044-4
[arXiv:hep-th/9908001 [hep-th]].

\bibitem{Aharony:2003sx}
O.~Aharony, J.~Marsano, S.~Minwalla, K.~Papadodimas and M.~Van Raamsdonk, ``The Hagedorn - deconfinement phase transition in weakly coupled large N gauge theories,''
Adv. Theor. Math. Phys. \textbf{8} (2004), 603-696 doi:10.4310/ATMP.2004.v8.n4.a1 [arXiv:hep-th/0310285 [hep-th]].

\bibitem{Shenker:2011zf}
S.~H.~Shenker and X.~Yin, ``Vector Models in the Singlet Sector at Finite Temperature,'' [arXiv:1109.3519 [hep-th]].

\bibitem{Kristensson:2020nly}
A.~T.~Kristensson and M.~Wilhelm, ``From Hagedorn to Lee-Yang: partition functions of $ \mathcal{N} $ = 4 SYM theory at finite N,'' JHEP \textbf{10} (2020), 006 doi:10.1007/JHEP10(2020)006 [arXiv:2005.06480 [hep-th]].

\bibitem{Hanada:2020uvt}
M.~Hanada, H.~Shimada and N.~Wintergerst, ``Color confinement and Bose-Einstein condensation,'' JHEP \textbf{08} (2021), 039 doi:10.1007/JHEP08(2021)039
[arXiv:2001.10459 [hep-th]].

\bibitem{deMelloKoch:2025rkw}
R.~de Mello Koch and A.~Jevicki, ``Hilbert Space of Finite $N$ Multi-matrix Models,'' [arXiv:2508.11986 [hep-th]].

\bibitem{Maldacena:1997re}
J.~M.~Maldacena, ``The Large N limit of superconformal field theories and supergravity,'' Adv. Theor. Math. Phys. \textbf{2} (1998), 231-252 doi:10.1023/A:1026654312961
[arXiv:hep-th/9711200 [hep-th]].

\bibitem{Gubser:1998bc}
S.~S.~Gubser, I.~R.~Klebanov and A.~M.~Polyakov, ``Gauge theory correlators from noncritical string theory,'' Phys. Lett. B \textbf{428} (1998), 105-114 doi:10.1016/S0370-2693(98)00377-3 [arXiv:hep-th/9802109 [hep-th]].

\bibitem{Witten:1998qj}
E.~Witten, ``Anti-de Sitter space and holography,'' Adv. Theor. Math. Phys. \textbf{2}, 253-291 (1998) doi:10.4310/ATMP.1998.v2.n2.a2 [arXiv:hep-th/9802150 [hep-th]].

\bibitem{JS1}
A.~Jevicki and B.~Sakita, ``The Quantum Collective Field Method and Its Application to the Planar Limit,''
Nucl. Phys. B \textbf{165} (1980), 511 doi:10.1016/0550-3213(80)90046-2

\bibitem{Jevicki:1991yi}
A.~Jevicki, ``Nonperturbative collective field theory,'' Nucl. Phys. B \textbf{376} (1992), 75-98
doi:10.1016/0550-3213(92)90068-M

\bibitem{Domokos}
Domokos, M., 2006. Multisymmetric syzygies. arXiv:math/0602303.

\bibitem{Procesi2}
C.~Procesi, ``A formal inverse to the Cayley-Hamilton theorem,'' Journal of algebra, (1987) 107(1), pp.63-74.

\bibitem{PI}
V. Drensky and E. Formanek, ``Quantitative Approach to PI-algebras,'' In Polynomial Identity Rings (pp. 19-35). Basel: Birkhäuser Basel (2004).

\bibitem{Pr}
C.~Procesi, \emph{The invariant theory of {$n\times n$} matrices}, Advances in Math. \textbf{19} (1976), no.~3, 306--381.

\bibitem{Sturmfels}
Bernd Sturmfels, “Algorithms in invariant theory,” Springer Science \& Business Media,
(2008).

\bibitem{hrtheorem}
M. Hochster, and J.L. Roberts, ``Rings of invariants of reductive groups acting on regular rings are Cohen-Macaulay,'' Advances in Mathematics, 13(2), pp.115-175 (1974).

\bibitem{bimahonian}
H. Barcelo, V. Reiner, and D. Stanton, ``Bimahonian distributions," Journal of the London Mathematical Society 77 no. 3 (2008): 627-646.

\bibitem{Rosele}
D.P. Roselle, ``Coefficients associated with the expansion of certain products." Proceedings of the American Mathematical Society 45 no. 1 (1974): 144-150.

\bibitem{Teranishi}
Y. Teranishi, “The ring of invariants of matrices,” Nagoya Math. J. 104 (1986) 149–161.

\bibitem{Harmark:2014mpa}
T.~Harmark and M.~Orselli, ``Spin Matrix Theory: A quantum mechanical model of the AdS/CFT correspondence,''
JHEP \textbf{11} (2014), 134 doi:10.1007/JHEP11(2014)134 [arXiv:1409.4417 [hep-th]].

\bibitem{McGady:2017rzv}
D.~A.~McGady, ``Temperature-reflection I: field theory, ensembles, and interactions,'' [arXiv:1711.07536 [hep-th]].

\bibitem{McGady:2018rmo}
D.~A.~McGady, ``Temperature-reflection II: Modular Invariance and T-reflection,'' [arXiv:1806.09873 [hep-th]].

\bibitem{king}
King, S.A., ``Fast Computation of Secondary Invariants,'' arXiv preprint math/0701270.



\end{thebibliography}
\end{document}